# Band Structure Engineering, Optical, Transport, and Photocatalytic Properties of Pristine and Doped $Nb_3O_7(OH)$: A Systematic DFT Study

Wilayat Khan[1], Alishba Tariq[1], Jan Minar[2], Sawera Durrani[1], Abdul Raziq[1], Sikander Azam[3], Khalid Saeed[4]

[1]Department of Physics, Bacha Khan University Charsadda, KP, Pakistan

[2]New Technologies-Research Centre, University of West Bohemia, Univerzitn´ı 8, 30100 Pilsen, Czech Republic

[3]Department of Physics, Riphah International University Islamabad, Pakistan

[4]Department of Chemistry, Bacha Khan University Charsadda, KP, Pakistan

**Abstract**

$Nb_3O_2(OH)$ has emerged as a highly attractive photocatalyst based on its chemical stability, energetic band positions, and large active lattice sites. Compared to other various photocatalytic semiconductors, it can be synthesized easily. This study presents a systematic analysis of pristine and doped $Nb_3O_7(OH)$ based on recent developments in related research. The current study summarizes the modeling approach and computationally used techniques for doped $Nb_3O_7(OH)$ based photocatalysts, focusing on their structural properties, defects engineering, and band structure engineering.

This study demonstrates that the Trans-Blaha modified Becke Johnson approximation (TB-mBJ) is an effective approach for optoelectronic properties of pristine and Ta/Sb-doped $Nb_3O_7(OH)$. The generalized gradient approximation is used for structure optimization of all systems studied. Spin-orbit (SO) coupling is also applied to deal with the Ta f orbital and Sb d orbital in the Ta/Sb-doped systems. Doping shifts the energetic band positions and relocates the Fermi level i.e. both the valence band maximum and the conduction band minimum are relocated, decreasing the band gap from 1.7 eV (pristine), to 1.266 eV (Ta-doped)/1.203 eV (Sb-doped). The band structures of pristine and doped systems reflect direct band behavior.  Investigation of the partial density of states reveals that O p orbital and Nb d/Ta d/Sb-d orbitals contributed to the valence and conduction bands, respectively. Optical properties like real and imaginary components of the dielectric function, reflectivity, and electron energy loss function are calculated by the OPTIC program implemented in the WIEN2k code. Moreover, doped systems shift the optical threshold to the visible region. Transport properties like effective mass and electrical conductivity are calculated, reflecting that the mobility of charge carriers increases with the doping of Ta/Sb atoms.

The reduction in the band gap and red-shift in the optical properties of the Ta/Sb-doped $Nb_3O_7(OH)$ to the visible region suggest their promising potential for photocatalytic activity and photoelectrochemical solar cells.

Corresponding Author: walayat76@gmail.com

## 1. Introduction

The global energy crises, with the expansion of industry and agriculture, especially associated with nonrenewable fossil fuels and environmental pollution, have become a major obstacle to sustainable development. Consequently, there is an increased interest in creating green technologies to address these crises. In investigating various green technologies, semiconducting photocatalytic materials have extensive interest due to their vast potential in both energy/environmental appliances. Many semiconducting materials like $TiO_2$, ZnO, $Fe_2O_3$, etc. have been synthesized for photocatalysis i.e. water splitting and degradation of organic pollutants, mainly depending on their band gaps [1-8].

$TiO_2$ is the most commonly used semiconductor among them, having been utilized by Fujishima and Honda as an electrode for water splitting [9]. Various modifications have been made i.e. doping, inorganic/organic molecular support, and nanostructuring, to improve the photocatalytic reaction. However, due to its large band gap, it utilizes only 4% of radiation in the UV region. This reduces the efficiency of $TiO_2$ and restricts its use for photocatalytic activity [10]. In addition to the above-mentioned semiconducting materials, organic-polymer nanomaterials like chitosan and cellulose are considered promising catalysts. They have paid tremendous attention due to abundant surface functional groups and environmental friendliness that improve the catalytic activity [11-13]. Recent reviews highlight the potential of inorganic and organic polymer-based nanomaterials, focusing on the advancement and applications of these organic-polymer-based composites, and inorganic nanomaterials. However, the limitation in the active lattice sites restricted the conventional inorganic-metal semiconductors and led to a discourse on the problems linked to metallic contamination [14, 15]. While organic polymers can be optimized flexibly, they encounter less stability and the challenges of separating catalysts from the products. When considering factors like catalyst optimization, selectivity, catalyst activity, stability, and the preparation of an effective photocatalytic system still faces major difficulties, especially concerning the cost and environmental sustainability.

The water-splitting reaction mechanism occurs at the surface of the semiconductor, and its high surface area is advantageous. The diffusion of the photo-generated charge carriers happens at the nanostructure's surface, which is countered by the recombination process, which decreases efficiencies [16]. The probability of electron-hole recombination can be reduced by small numbers of defects in the lattice and with small dimensions of nanostructures. The large surface area enhances the active sites for reaction [17]. Like Graphene and its various doped configurations with N atoms, low-dimensional materials have greater potential in device applications [18]. M. A. Dar [19] investigated the photocatalytic activity of highly stable $B_2$-doped g-$C_2N$ and g-$C_6N_6$ monolayers with various dimensions leading to the reduction of Urea and Ammonia. It has also been reported that the incorporation of a single impurity into g-$C_2N$ and GaN monolayers greatly impacts the band gap, making the parent material favorable for electrochemical $CO_2$RR [20, 21]. Therefore, a 3D (three-dimensional) hierarchical $Nb_3O_7(OH)$ superstructure was prepared via a step template-free hydrothermal strategy [22]. This material was further studied theoretically to investigate the energetic band position of this system and confirmed that this system has higher mobility than $Nb_2O_5$, which results in high photocatalytic activity [23]. Further, Betzler [24] reported the change in morphology and found an enhancement in the light-induced water-splitting reaction through Ti doping into the 3D $Nb_3O_7(OH)$ nanostructure.

This study presents a density functional functional (DFT) understanding of the structural, electronic, optical, and thermoelectric properties of the pristine and doped $Nb_3O_7(OH)$. Tantalum (Ta) and Antimony (Sb) are employed as dopants to investigate the influence of metal incorporation. Detailed investigations like structural, electronic band structure, partial density of states (PDOS), real/imaginary part of the dielectric function, and transport properties including the effective mass and thermoelectrical conductivity for the entire systems. It also includes the photocatalytic activity of the pristine and doped systems. The impact of the positioning of the dopants in the crystal lattice is also studied. The results offer a comprehensive analysis of the influence of metal doping on $Nb_3O_7(OH)$ properties.

## 2. Methodology

To understand the doping effect, a $2 \times 2 \times 1$ supercell of $Nb_3O_7(OH)$ was constructed, where one Nb atom was substituted by one Tantalum (Ta) atom (Ta:$Nb_3O_7(OH)$). Similarly, one Antimony (Sb) atom replaced one Nb atom at the same lattice site in the crystal (see Fig. 1).

All electronic structure calculations were conducted using plane DFT based on the Generalized Gradient approximation (GGA) [25-27]. This approach was implemented in the WIEN2k code [28], which utilizes the full-potential plane wave method (FP-LAPW) [29]. The GGA approximation is one of the most precise functional for geometric optimization of the crystal. We used it for all crystal systems. The muffin tin sphere's radii are selected as 1.65 (Niobium Nb), 1.73 (Ta/Sb), 1.25 (Oxygen O), and 0.67 (Hydrogen H), respectively in all three crystals.

The energy cut-off/plane wave cut-off is settled up to 7.0. The spherical harmonics were expanded inside the muffin tin spheres up to $l_{max}=10$, while the charge density (expanded in Fourier series) was carried up to $G_{max}= 12$ $(a.u)^{-1}$. To obtain the converged energy, a Monkhorst-Pack k-mesh of $1 \times 2 \times 2$ was selected for Tantalum(Ta)/Antimony(Sb) in the first irreducible Brillouin zone, while a denser k-mesh of $3 \times 9 \times 9$ was selected for self-consistent calculations [30].

In this study, Trans-Blaha modified Becke-Johnson approximation plus spin-orbit coupling (TB-mBJ+SO) is used to calculate the electronic structure and optical properties calculation of doped systems. In addition, the TB-mBJ approximation is more often used to predict accurate band gaps than the GGA approximation [31, 32]. To perform these calculations, we build three structures: (1) the pristine crystal of $Nb_3O_7(OH)$ comprises 24 atoms. A $2 \times 2 \times 1$ supercell of $Nb_3O_7(OH)$, consisting of 96 atoms in total, was then built to study the impact of the dopant. The last two structures are mainly assigned to the Ta/Sb-doped $Nb_3O_7(OH)$. There are a total of 23 atoms of Nb after substituting one Nb atom which sits in the corner-sharing octahedra by Ta atom (Ta:$Nb_3O_7(OH)$) and Sb atom (Sb:$Nb_3O_7(OH)$), and kept the concentration of the Ta and Sb atoms up to 4.16 %, respectively.

Optical properties were calculated from the OPTIC code used in the WIEN2k simulation code [33]. The energy loss function for the studied compounds was obtained using Fermi's golden rule and dipole matrix element. The BoltzTraP code [34], an interfaced code in the WIEN2k code, is used to find the thermoelectrical properties based on Boltzmann's semi-classical theory.

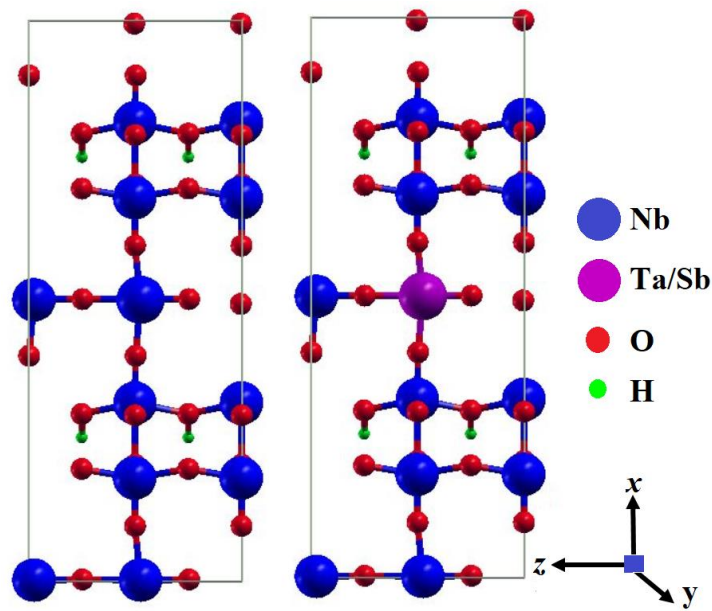

**Figure 1.** Schematic presentation of the pristine and Ta/Sb-doped $Nb_3O_7(OH)$

## 3. Results and Discussion:

### 3.1. Structural Properties

We performed internal/external optimization of the parent and Sb/Ta-doped $Nb_3O_7(OH)$ through the generalized gradient approximation (GGA). The optimized lattice parameters of the studied systems are listed in Table 1 and found that there is very minor variation in the lattice parameters while moving from parent to Sb/Ta-doped $Nb_3O_7(OH)$ to see the impact of these dopants on the bond lengths. We also listed the internal parameters (bond lengths) in Table 2., along with x/y/z-axes. Comparing these parameters along different directions for Nb-O, Sb-O, and Ta-O, it was found that there is more increase along the y-axis than x/z-axes. It means that there is an increase in the surface area of the doped systems, which increases in active reactive sites. The studied systems also have higher catalytic activity like other 2D (two-dimensional) monolayers having high surface area, and distinctive properties suitable for catalytic activity [35].

Table 1. Optimized lattice parameters of parent and doped $Nb_3O_7(OH)$

| Compound Name | Optimized Lattice Parameters in Å | | | |
|---|---|---|---|---|
| $Nb_3O_7(OH)$ | a | 20.797 | 7.663 | 7.895 |
| $Sb:Nb_3O_7(OH)$ | b | 20.814 | 7.680 | 7.912 |
| $Ta:Nb_3O_7(OH)$ | c | 20.830 | 7.697 | 7.929 |

Table 2. Optimized bond lengths between Nb-O, Sb-O and Ta-O of parent and doped $Nb_3O_7(OH)$

| Compound Name | Crystallographic direction in Å | | | | |
|---|---|---|---|---|---|
| Nb3O7(OH) | Nb-O | Along x-axis | 1.917 | 1.929 | 1.932 |
| Sb:Nb3O7(OH) | Sb-O | Along y-axis | 1.799 | 1.996 | 2.116 |
| Ta:Nb3O7(OH) | Ta-O | Along z-axis | 2.018 | 2.017 | 2.010 |

## 3.2. Electronic Band Structure

The band structure explains the behavior of the electrons in a material by delineating their energy and momentum states. This study performed electronic structure calculations in a large supercell (pristine/doped-$Nb_3O_7(OH)$). The dopant level can be verified from the band structure and density of states of the investigated systems (see Fig. 2). Investigating the material's band gap is essential for understanding the charge carriers/optical properties. Note that the ideal band gap lies in the range of 1.1 to 1.4 eV for photovoltaic applications, and for photocatalysis, the optimal band gap lies in the range of 1.8 to 2.2 eV. Firstly, it is widely established that GGA functional underestimates the band gap more than what is observed experimentally. In contrast, TB-mBJ approximation typically yields band gaps that equate or exceed the experimentally observed value. Figs. 2(a-c) shows the electronic band structures of pristine and doped $Nb_3O_7(OH)$. The Fermi energy level (marked by a dashed line) is set to zero energy. The valence band maximum (VBM) and the conduction band minimum (CBM) are located at the Γ point, showing that the systems are direct-band semiconductors. The energy width between VB and CB (energy band gap) for the pristine system is 1.7 eV/3.1 eV (optical band gap) is similar to the band gap value reported by Khan et. al., [24] using TB-mBJ approximation. Upon the analysis of Fig. 2, it is clear that all systems produced similar band structures, having both the valence band maximum VBM and the conduction band minimum CBM positioned at the Γ point.

The fundamental band gaps of the Ta/Sb-doped $Nb_3O_7(OH)$ are 1.266/1.023 eV, where the optical band gaps are 2.59 eV and 2.30 eV, respectively, as shown in Fig. 2(b & c). In both systems, the CBM (conduction band minima) is shifted downward, while the Fermi energy level moves closer to the conduction band, indicating the n-type semiconductor behavior. This can be attributed to the larger valance electrons in Ta and Sb atoms than in the Nb atom. At the top of the VBM, the band splitting is caused by spin-orbit (SO) coupling employed to Sb d orbitals and Ta d orbitals, respectively. The band gap values for the entire Ta/Sb:$Nb_3O_7(OH)$ systems were observed alike irrespective of the various positions of the dopant in the crystal lattice of the pristine crystals. The consistency in the band gap values does not depend on the dopant position. The conduction/valence bands depict greater dispersion features with relevant symmetry points suggesting a small effective mass of the holes/electrons in the entire system, reflecting interesting electronic characteristics. It is important to mention that the investigated compound's band gap values make them the most suitable candidate for photo-electrochemical (PEC) applications.

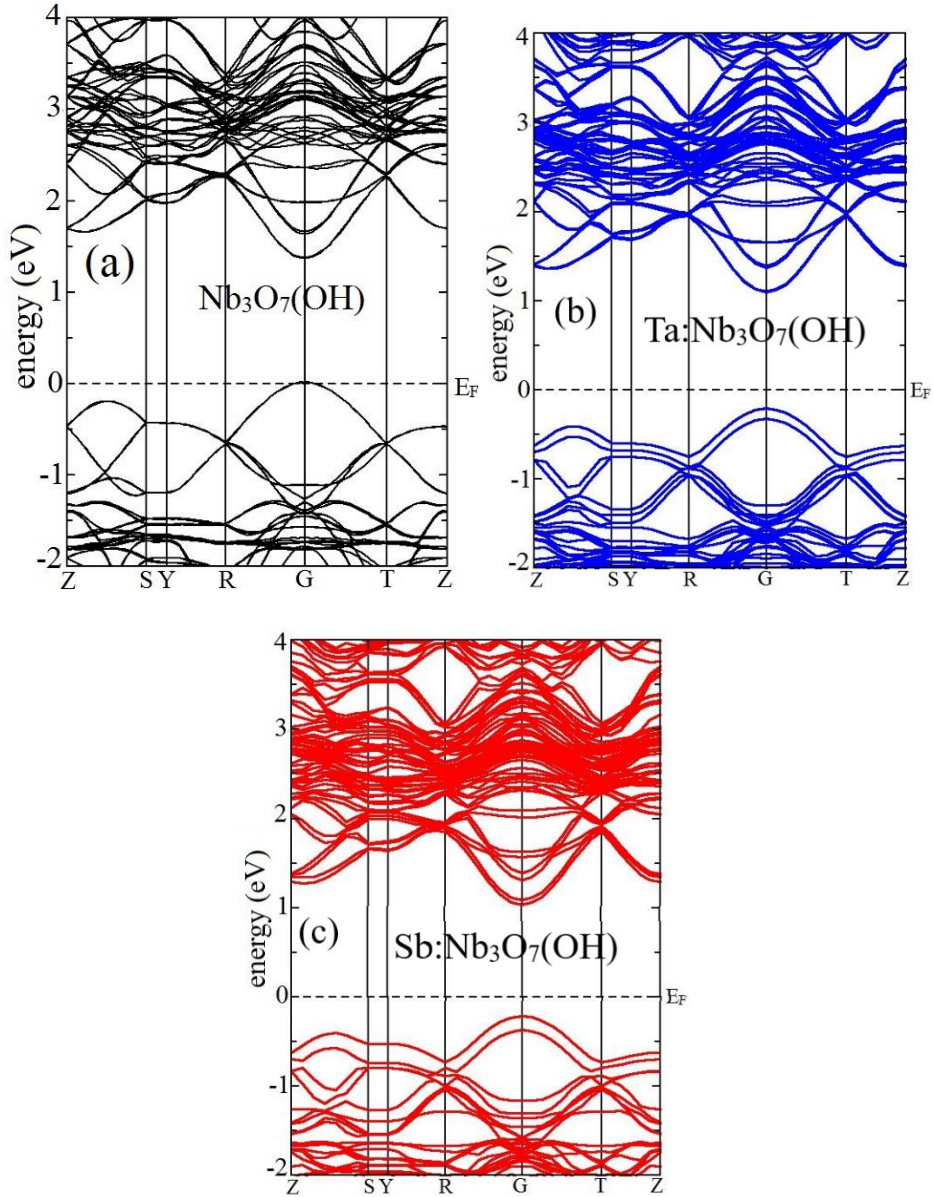

**Figure 2.** Schematic description of calculated electronic band structures. (a) $Nb_3O_7(OH)$, (b) $Ta:Nb_3O_7(OH)$, and (c) $Sb:Nb_3O_7(OH)$, using Trans-Blaha modified Becke Johnson approximation (TB-mBJ/TB-mBJ+SO).

### 3.3. Density of States

For the detailed analysis of the electronic structure properties of the pristine and doped $Nb_3O_7(OH)$, the partial density of states was simulated using TB-mBJ/TB-mBJ+SO approximation for the parent/doped systems. Figs. 3(a-d) depict the partial density of states (PDOS) of the $Nb_3O_7(OH)$, and Ta/Sb-doped $Nb_3O_7(OH)$, respectively.

Figs. 3(a-d) illustrates that, in a pristine system, the region around the Fermi level is mainly composed of O p orbitals. In contrast, CBM is primarily made up of Nb d orbitals. In a region from -2.0 to -4.0 eV, the O p orbitals and Nb d orbitals depict strong hybridization and the Nb/O s orbitals show strong hybridization too. The conduction band (CB) is predominantly made up of Nb d orbitals along with the minor contribution of other orbitals of other atoms, respectively.

In the Ta-doped system, the position of the CBM is expected to alter with minor modification in the VBM due to the introduction of 4d metals like Sb and 5d metals like Ta in the crystal lattice of the $Nb_3O_7(OH)$. Unlike the impurities bands at the top of the VBM, strong consistencies in the behavior of the doped systems have been observed near CBM. A strong perturbation of the CBM/VBM is due to the addition of Sb and Ta atoms, however, the Sb atom contributes more clearly to impurity bands than the Ta atom (see Fig. 2 and Fig. 3). Fig. 3 illustrates the contribution of the Ta d/f orbitals in addition to the O p orbitals around $E_F$, and the hybridization of Ta f orbital with Nb p orbitals around 1.0 eV. Whereas, the impurities (Ta/Sb atoms) shift the H atom to lower energy, and the lower part of the conduction band is predominantly constructed by Ta d orbital than the pristine and Sb-doped system, where the main contributor to CB is Nb d orbitals. The band gap values, 1.266 eV (Ta:$Nb_3O_7(OH)$)/1.023 eV (Sb:$Nb_3O_7(OH)$), are decreased, which can be explained based on the introduction of the 4d and 5f orbitals of the impurity atoms, which shifted up the Fermi level, downshift the CBM, respectively. Considering the effect of the dopant location, we found the same findings of the PDOS.

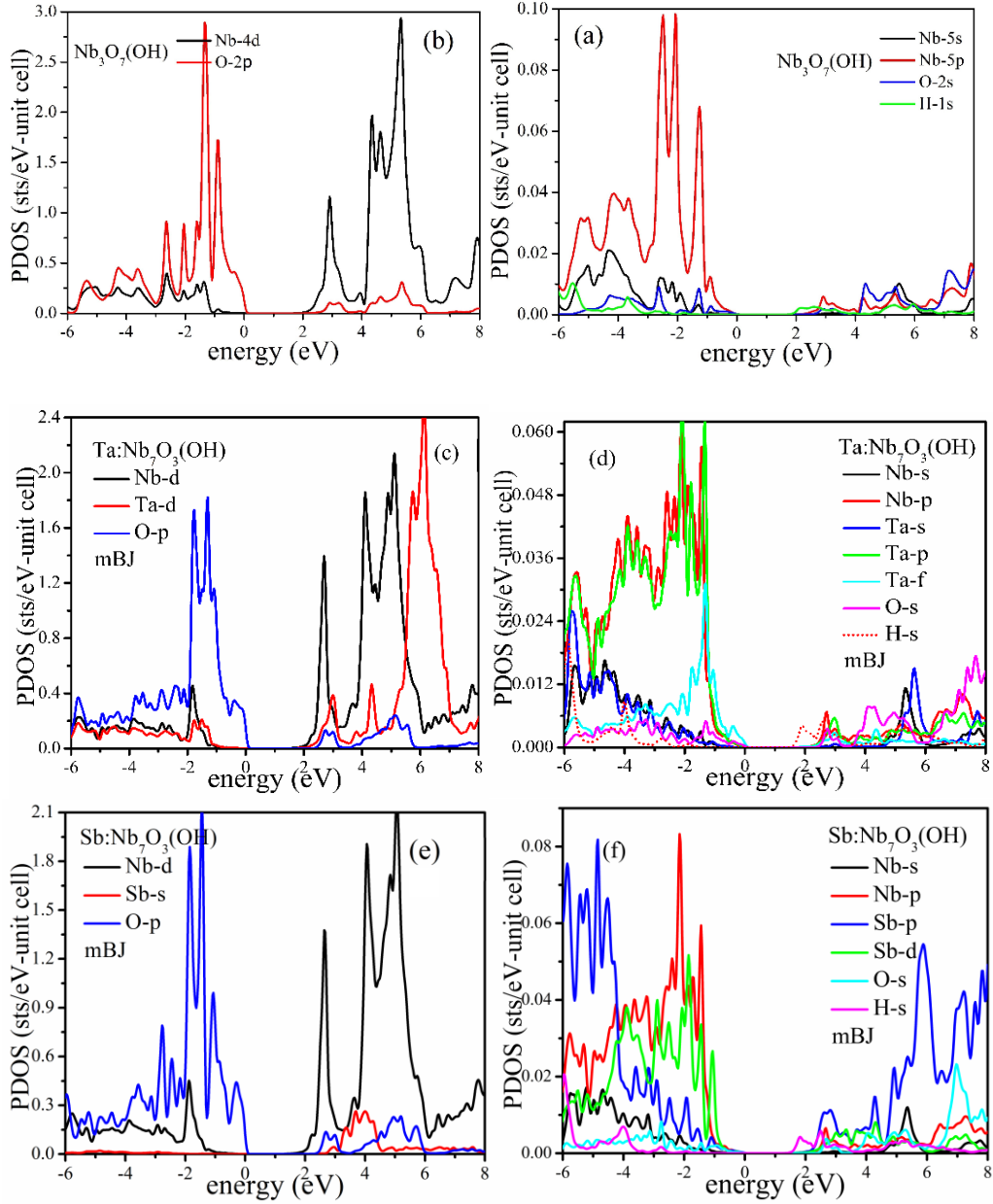

**Figure 3.** Calculated partial density of states (PDOS) using Trans-Blaha modified Becke Johnson approximation (TB-mBJ/TB-mBJ+SO). (a, b) PDOS of Nb$_3$O$_7$(OH). (c, d) PDOS of Ta:Nb$_3$O$_7$(OH) (e, f) PDOS of Sb: Nb$_3$O$_7$(OH).

## 3.4. Optical Properties

To gain deep insight and further analyze the investigated compounds, the linear optical response properties have been studied to clarify the effect of the dopant on Nb$_3$O$_7$(OH). It is important to note that semiconductors play a crucial role in device applications because their band gap defines the threshold energy and differentiates between the absorption and transparency regions.

The optical parameters of pristine and doped $Nb_3O_7(OH)$ were analyzed by complex dielectric function $\varepsilon(\omega)$. The calculation of the imaginary part ($\varepsilon_2(\omega)$) is based on the momentum matrix element, and is described by the following equation:

$$\varepsilon_2(\omega) = \frac{2\pi e^2}{\Omega \varepsilon_0} \sum_{k,v,c} |<\varphi_k^c|\hat{u}r|\varphi_k^v>|^2 \delta(E_k^v - E) \tag{1}$$

where $e$, $\Omega$, and $v$ referred to as electronic charge, unit cell volume, frequency of radiation. The induced electric field polarization and the wave function associated with conduction/valence at k are symbolized as $\hat{u}$, and $\varphi_k^c/\varphi_k^v$, respectively. The real part $\varepsilon(\omega)$ can be obtained *via* the Kramers-Kronig relation.

The real $\varepsilon_1(\omega)$ and imaginary $\varepsilon_2(\omega)$ parts of the dielectric function for pristine and doped-$Nb_3O_7(OH)$ are illustrated in Figs. 4(a, b). The introduction of the dopant (Sb and Ta atoms) in the crystal lattice of the $Nb_3O_7(OH)$ structure increases the dielectric constant in the following sequence 4.02 (pristine)/4.143 (Sb-doped)/4.149 (Ta-doped), respectively. At zero frequency, the doped crystals ($Nb_3O_7(OH)$) have a higher dielectric constant than pristine, which means the electronic polarizability for the doped systems is greater than that of the pure systems. Fig. 4a depicts that the peaks below the zero line (marked as a green dash line) define the near-metallic nature of the investigated crystals. The reduction in the negative value of $\varepsilon_1(\omega)$ could be explained as introducing the dopants i.e. 4d/5d metals (Sb, Ta atoms) in the crystal lattice of the studied systems.

Fig. 4b presents $\varepsilon_2(\omega)$ considered only electronic participation. The $\varepsilon_2(\omega)$ spectra of the studied systems are depicted in Fig. 4b indicating that the fraction of electromagnetic radiation is lost when it goes through materials. The TB-mBJ functional yields an optical band gap value of 3.1 eV for the pristine system, which agrees with the experimental data (3.1±0.1 eV) [22]. The calculated optical threshold is 3.25 eV for the undoped system, which is higher than the Ta-doped (2.4 eV) and then followed by the Sb-doped (2.11 eV) system, respectively (see Fig. 4b). Pristine $Nb_3O_7(OH)$ has an optical band gap value of 3.1 eV, which limits the absorption of solar radiation to only 4 % in the UV region as reported in previous studies [36, 37]. The major peaks of the $\varepsilon_2(\omega)$ describe the transition from occupied (VB) to the unoccupied orbitals (CB). Fig. 4b reveals three peaks for the pristine system, among them two are lying at 4.93 eV and 6.19 eV, and the third peak (broader one) is located at 8.88 eV with two shoulders, respectively. In a pristine system, these peaks are associated with the electronic transitions from O p orbital (occupied band) to Nb d orbital (unoccupied band) aligned with the results reported in Khan et. al., [24].

There is a shift in the optical threshold towards the visible region attributed to the dopant in the crystal lattice of the systems. The main peak of the Ta-doped system is positioned at 5.05 eV, whereas the main peak for Sb-doped is observed at the same energy. The only difference is the intensity variations. The transitions from O p orbital to Nb d orbitals are responsible for the main peaks in the spectra of the doped system, and the following peaks are made from the transition of O p orbital and Ta d orbital. Despite the main peaks, there is a shoulder in the spectra in the visible region that matches the incorporation of Sb/Ta atoms (see Fig. 3), and these are associated with the transition from O p orbital to H s orbital.

Fig. 4c depicts the reflectivity $R(\omega)$ of the investigated systems. The investigated systems possess two main peaks with some small peaks and shoulders. The maximum peaks are observed at 6.14 eV and 9.78 eV similar to the previously published data, which reflect 28 % (main peak) and 25 % (second peak) of the incident radiations. The increased calculated reflectivity of the investigated systems indicates their potential for use in photo-electrochemical (PEC) appliances.

Fig. 4d shows the dependence of electron energy loss function $L(\omega)$ upon the incident photon energy for pristine/doped $Nb_3O_7(OH)$. The EEL spectra come from the free oscillations of the valence band's electrons upon irradiation, which causes a plasma frequency. The $L(\omega)$ appears from the energy points in $\varepsilon_1(\omega)$, where $\varepsilon_1(\omega)$ crosses the zero line. It is obvious from Fig. 4d that the $L(\omega)$ spectra increase with photon energy up to 14.0 eV for the pristine system and up to 12.5 eV for doped systems. The positions of the peaks depend on the composition of the systems, and the heights of these peaks are due to plasmon resonance. The calculated EEL spectra for the doped systems have similar features with the pristine except for a shift towards lower energy.

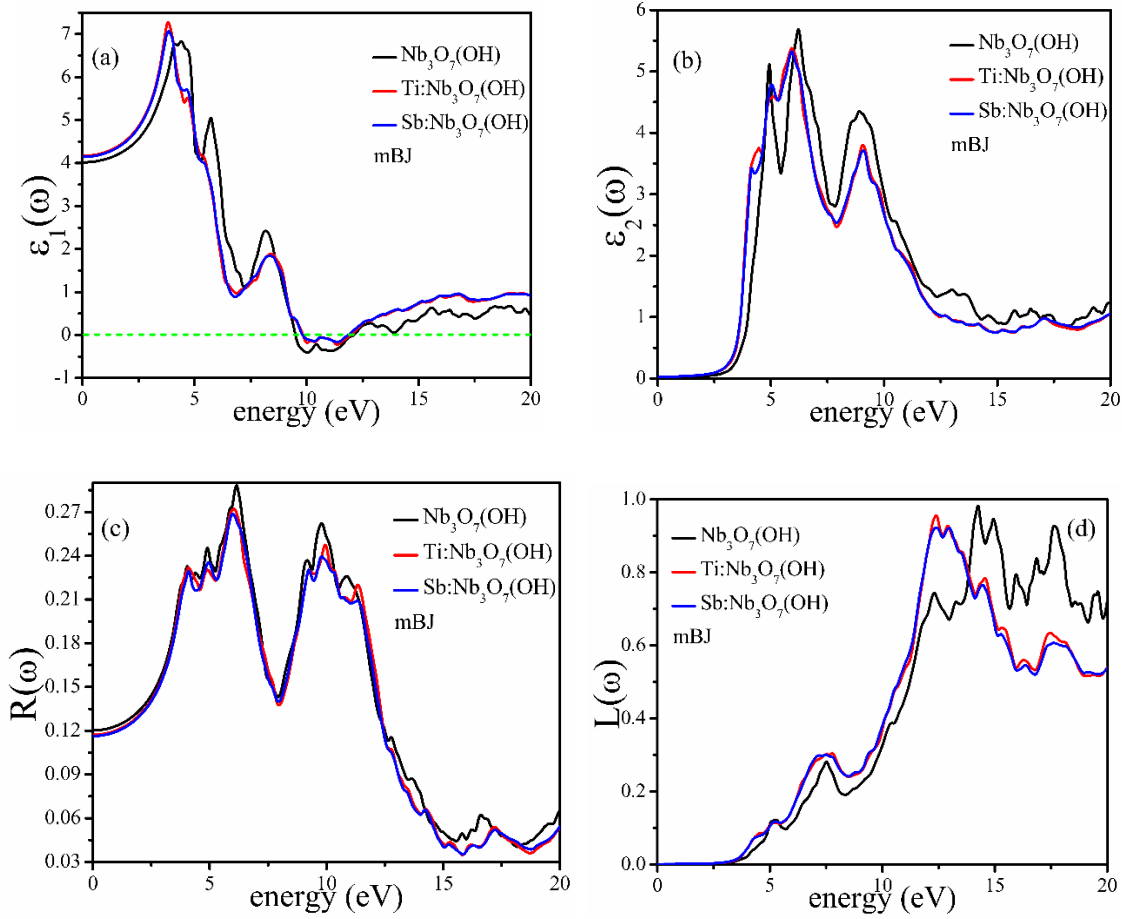

**Figure 4.** (a) Calculated real part of the dielectric function (b) Imaginary part of the dielectric function (c) Reflectivity, and (d) Electron energy loss function spectra (EELs) of the pristine and Ta/Sb-Nb$_3$O$_7$(OH), using Trans-Blaha modified Becke Johnson approximation

### 3.5. Transport Properties

In solid-state physics, effective mass is a crucial quantity influencing how charge carriers, such as electrons and holes, move through materials. It considers the periodic potential's effect when defining the apparent mass of a carrier moving through a crystal lattice. The effective mass of electrons and holes is associated with the curvature of the valence band (for holes) or conduction band (for electrons) around the band extremum. These masses match the masses that appear to participate in conduction properties. In addition, the anisotropy of charge carrier conduction might affect the effective mass, with $m^* < 0.5$ potentially having significant mobility. Furthermore, the binding energy of excitons is influenced by the effective mass $m^*$ i.e. lower $m^*$ corresponds to a lower binding energy ($E_b$).

The effective mass values for the holes and electrons for pristine and doped $Nb_3O_7(OH)$ are listed in Table 3. A decrease was found in the effective mass of electrons $m_e^*$ and holes $m_h^*$, while moving from pristine to doped systems i.e. 0.147 (pristine)→0.072 (Sb-doped)→0.064 (Ta-doped) for $m_e^*$ and it decreases from 0.351 (pristine)→0.04 (Sb-doped)→0.0344 (Ta-doped) for $m_h^*$, respectively. It implies that both doped systems have lower effective mass values.

Comparatively, Ta-doped Nb3O7(OH) depicts substantiationally reduced effective masses ($m_e^*/m_h^*$) than the other two systems This implies that the Ta-doped system has higher mobility, where the inverse relation of the effective mass and carrier mobility applies. It greatly influences the material's electronic properties such as conductivity, optical behavior, and possible applications in electronic and optoelectronic fields. The lowered value of effective masses for both carriers demonstrates that the Ta-doped $NH_3O_7(OH)$ may have superior charge transport properties, making a material more potential for electrical and optoelectronic.

**Table 3:** Calculated effective masses of pristine and Ta/Sb-doped $NH_3O_7(OH)$.

| Materials | $m_e^*$ | $m_h^*$ |
|---|---|---|
| $Nb_3O_7(OH)$ | 0.147 | 0.351 |
| $Ta:Nb_3O_7(OH)$ | 0.064 | 0.0344 |
| $Sb:Nb_3O_7(OH)$ | 0.072 | 0.04 |

To further elucidate the prospect of pristine and doped $Nb_3O_7(OH)$ in thermoelectric appliances, thermoelectrical conductivity was simulated using the BoltzTraP code. Fig. 5 illustrates electrical conductivity $\sigma^{ave}(\mu, T)$ at 300 K for all systems *vs* chemical potential, which can be determined from the density of states (DOS). The region below and above the Fermi level is designated as the n-type region (electron region) and the p-type region (hole region). The $\sigma^{ave}(\mu, T)$ in the n-type region for all systems is higher than the p-type region, whereas the Ta-doped system depicts a higher value in the lower energy region. It is important to note that, Ta doping can also result in new scattering centers, which might reduce overall $\sigma^{ave}(\mu, T)$ and possibly reduce some mobility advantages through greater scattering. Unlikely Ta-doped $Nb_3O_7$(OH), Sb-doped $Nb_3O_7$(OH) has a more complex band structure with several peaks in the DOS, which may restrict carrier movement because of probable localization effects.

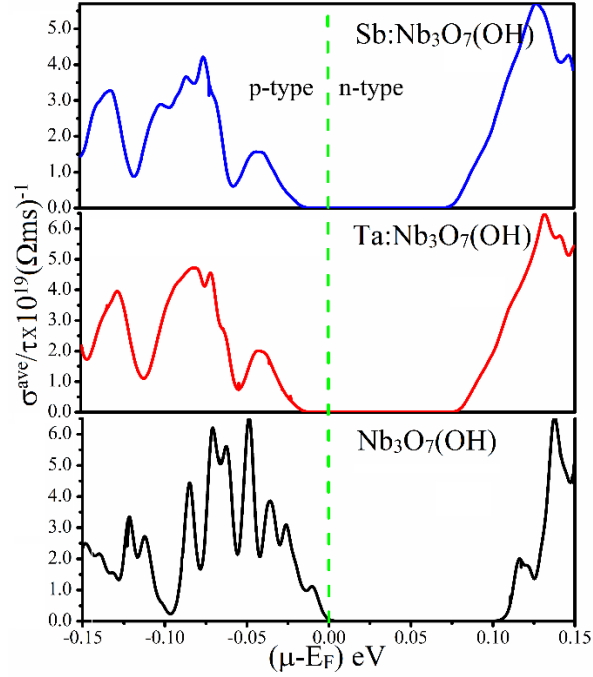

**Figure 5.** Calculated electrical conductivity at 300 K using constant relaxation time approximation employed in BoltzTraP code.

### 3.6. Determining the conduction band and valence band positions

Knowing the ionization energy and the electron affinity of the constituent element of the investigated systems allows us to determine the positions of the conduction/valence band edges in solid aqueous solution composition. These positions can be evaluated for the pristine and Sb/Ta-doped $Nb_3O_7(OH)$, using the following formulae [38, 30]:

$$E_{CB}^0 = E_e - \chi + 0.5E_g, \text{ and } E_{VB}^0 = E_{CB} - E_g, \tag{2}$$

where $\chi = \{(\chi_{Nb})^m \times (\chi_O)^n \times (\chi_H)^P\}^{1/m+n+P}$ (for pristine), (3)

$\chi = \{(\chi_{X=Ta/Sb})^m \times (\chi_{Nb})^n \times (\chi_O)^p \times (\chi_H)^q\}^{1/m+n+P+q}$ (for the doped system) (4)

and

$\chi_X = 1/2[E_{EA}^X + E_{ion}^X]$, where X=Nb, O, H, and Ta/Sb (5)

Where $\chi/E_{CB}^0$ is the electronegativity, the energy of the conduction band edge, and $E_e$ is a constant equal to 4.5 eV [39], and it is based on the composition of the element. The ionization energy and electron affinity energy of the Nb/O/H/Ta/Sb elements are symbolized by $E_{EA}^X/E_{ion}^X$, where the optical band gap of the pristine/doped Nb$_3$O$_7$(OH) is denoted by $E_g$. Electron affinity is the energy gained by moving electrons outside the semiconductor from the vacuum to the bottom of the CB inside the semiconductor. The computed values of the ionization energy, electronegativity, and the positions of both the CB/VB of pristine/doped-Nb$_3$O$_7$(OH) are listed in Table 4. The substitution of the Ta/Sb in the crystal lattice of the Nb$_3$O$_7$(OH) system showed that the VB edge and CB edge shifted small positive/negative potential, respectively. These findings show that the doped compounds show better photocatalysis than the pristine system.

**Table 4:** Calculated ionization energy of atoms (KJ/mole) Electron affinity (eV), and conduction/valence band edge position of pristine and Ta/Sb-doped NH$_3$O$_7$(OH).

| Name of the element/Compound | $E_{ion}$ in kJ/mole | $\chi$ in eV | $E_{CB}$ in eV | $E_{VB}$ in eV |
|---|---|---|---|---|
| Nb | 652 | | | |
| Ta | 760 | | | |
| Sb | 834 | | | |
| H | 1312 | | | |
| O | 1314 | | | |
| Nb$_3$O$_7$(OH) | | 520.15 | 3.17 | 0.07 |
| Ta:Nb$_3$O$_7$(OH) | | 537.34 | 2.95 | 0.36 |
| Sb:Nb$_3$O$_7$(OH) | | 537.32 | 2.85 | 0.55 |

**Conclusion**

It is concluded that fundamental properties like band gap, dielectric constant, effective mass, electrical conductivity, and photocatalytic properties are important for the semiconductors to use potentially in photovoltaic and photochemical appliances.

Theoretical investigation of the electronic structure, band gap engineering, optical/thermoelectric properties, and photocatalytic properties influenced by the dopants using the full potential linearized augmented plane wave method employed in the WIEN2k program. The generalized gradient approximation (GGA) was used for internal geometry optimization. Trans Blaha modified

Becke Johnson approximation (TB-mBJ) was utilized to determine the accurate band gap. For pristine $Nb_3O_7(OH)$, the calculated finding specified a band gap of 1.7 eV (fundamental band gap)/3.1 eV (optical band gap), which is aligned with the experimental data. The SO coupling was applied to doped systems. The substitution of Ta and Sb atoms in the crystal lattice of $Nb_3O_7(OH)$ resulted in a shift in the band gap and repositioning of the Fermi level. The VBM led to a downward shift, and CBM shifted towards the Fermi level, resulting in the band gap i.e. 1.7 eV→ 1.266 eV→1.203 eV. The calculated PDOS of the pristine system described that the O p orbitals mainly contributed to VB, and the Nb d orbital constructed the CB. While in the doped $Nb_3O_7(OH)$, the CB is made of Nb d orbitals and Ta d orbitals, and the O p orbitals still dominated the VB in addition to the small contribution of the orbitals from doped atoms (Ta-doped $Nb_3O_7(OH)$). In Sb-doped $Nb_3O_7(OH)$, the CB is constructed by Nb d/Sb s orbitals, respectively. Further, the dopants induced a red shift in the optical threshold to the visible region. Furthermore, the doped systems showed larger mobility and photocatalytic activity than the pristine system. These findings confirmed that Ta/Sb-doped $Nb_3O_7(OH)$ holds the potential to be an outstanding candidate for photoelectrochemical solar cells.

## Acknowledgments

J. M. would like to thank the QM4ST project with Reg. No. CZ.02.01.01/00/22_008/0004572, co-funded by the ERDF as part of the MˇSMT.